\documentclass[twocolumn,showpacs,prb,preprintnumbers,superscriptaddress,amsmath,amssymb]{revtex4}
\usepackage{graphicx}
\usepackage{dcolumn}
\usepackage{bm}
\usepackage{amsmath,amssymb}
\newcommand{\be}{\begin{eqnarray}}
\newcommand{\ee}{\end{eqnarray}}

\begin{document}
\title{Characterization of topological phases of modified dimerized Kitaev chain via edge correlation functions}
\author{Yucheng Wang}
\affiliation{Beijing National
Laboratory for Condensed Matter Physics, Institute of Physics,
Chinese Academy of Sciences, Beijing 100190, China}
\affiliation{School of Physical Sciences, University of Chinese Academy of Sciences, Beijing, 100049, China}
\author{Jian-Jian Miao}
\affiliation{Kavli Institute for Theoretical Sciences,
Chinese Academy of Sciences, University of Chinese Academy of Sciences, Beijing, 100190, China}
\author{Shu Chen}
\thanks{schen@iphy.ac.cn}
\affiliation{Beijing National
Laboratory for Condensed Matter Physics, Institute of Physics,
Chinese Academy of Sciences, Beijing 100190, China}
\affiliation{School of Physical Sciences, University of Chinese Academy of Sciences, Beijing, 100049, China}
\affiliation{Collaborative Innovation Center of Quantum Matter, Beijing, China}
\begin{abstract}
We study analytically a modified dimerized Kitaev chain model under both periodic and open boundary conditions and determine its phase diagram by calculating the topological number and edge correlation functions of Majorana fermions. We explicitly give analytical solutions for eigenstates of the model under open boundary conditions, which permit us to calculate  two introduced edge correlation functions of Majorana fermions analytically. Our results indicate that the two edge correlation functions can be used to characterize the trivial, Su-Schrieffer-Heeger-like topological and topological superconductor phases. Furthermore, we find the existence of two different coupling patterns of Majorana fermions in topological superconductor phase characterized by different edge correlation functions, which can also be distinguished by investigating the survival probability of an edge Majorana under sudden quenching of parameters.
\end{abstract}
\maketitle
\section{Introduction}
Tremendous advances in studying the topological insulators and topological superconductors \cite{Zhang} have been seen in the last decade, which have shed light on our understanding of the topological properties of solids \cite{Ludwig,Ryu,Tanaka}. According to the topological band theory,  each topological class  can be characterized by a topological invariant associated with energy eigenstates in momentum spaces, which cannot change except that the closing-reopening process of the systematic bulk gap occurs. The topological properties of a system can be unveiled by calculating the topological indices in momentum space or analysing the edge states in topological insulator or Majorana zero modes in topological superconductors under open boundary conditions (OBC). Although the edge states can be generally studied by numerically diagonalizing the Hamiltonian under OBC, it is not a easy task for analytically solving the edge states and calculating the edge correlation functions, which may deepen our understanding of the topological edge states and provide useful informations for the finite-size effects.

As a prototype model of one-dimensional (1D) topological p-wave superconductor with edge Majorana modes \cite{kitaev}, the Kitaev chain model has attracted intensive studies. This is not only due to the fundamental interest in exploring Majorana fermions (MFs) and this model can be realized in some experiments \cite{Kane,Alicea,Sarma2,Fujimoto,Sarma3,Mourik}, but also some extended Kitaev models, such as Kitaev models with periodic or quasiperiodic modulation of chemical potential \cite{Lang,Chen2,Sen,Hu,Wang,Loss}, the dimerized Kitaev model \cite{Nagaosa,Chen,Sugimoto} and some quasi-1D Kitaev models \cite{Zhou,Kells,Wakatsuki}, exhibit rich physical properties. In this work, we shall study a modified dimerized Kitaev model with the hopping strength and the superconductor pairing strength being modified alternatively. In the absence of paring terms or dimerization terms, our model is able to  reduce to the Su-Schrieffer-Heeger (SSH) model \cite{ssh,ssh2} or Kitaev chain model \cite{kitaev}, respectively, however our model has no a completely dimerized limit, in which the system is composed of separated dimerization cells, as the modulations of hopping and superconducting pairing terms are completely inverse.
By studying both cases with the periodic and OBC, we demonstrate that our model displays a rich phase diagram.
Corresponding to the two different boundary conditions, we determine the phase diagram  by using two different methods: either directly calculating the topological number or calculating two introduced edge correlation functions. As both of the them give the same phase diagram, the latter method furnishes a direct physical picture to understand the phase diagram and gives two different coupling patterns of MFs in the topological superconductor (TSC) phase characterized by different edge correlation functions.
To see the different coupling patterns more clearly, we further investigate the survival probability of an edge Majorana \cite{Patel,Rajak,Sacramento} under sudden quenching of parameters in the regions corresponding to the two different coupling patterns.

This paper is organized as follows: in Sec. \ref{model}, we introduce the modified dimerized Kitaev model and obtain its phase diagram by calculating the Majorana number analytically under periodic boundary conditions (PBC). In Sec. \ref{correlation}, we rewrite the Hamitonian in the Majorana fermion representation firstly  and study the model under OBC. We then exactly diagonalize the Hamiltonian by using the method of singular value decomposition (SVD) and calculate analytically two edge correlation functions of Majorana fermions, which gives rise to the phase diagram of the system.
In Sec. \ref{dynamics}, we discuss the survival probability of an edge Majorana under sudden quenching of parameters in the two different coupling cases of TSC phase. A brief summary is given in Sec. \ref{summary}.

\section{Model Hamiltonian and the phase diagram}
\label{model}
We consider a modified dimerized Kitaev superconductor model described by
\be
H &=& -t \sum_{j} [(1+\lambda)c^\dagger_{j,A}c_{j,B} + (1-\lambda)c^\dagger_{j,B}c_{j+1,A}+H.c.]\nonumber\\
  &-& \Delta \sum_{j}[(1-\lambda)c^{\dagger}_{j,A}c^{\dagger}_{j,B}+ (1+\lambda)c^\dagger_{j,B}c^\dagger_{j+1,A} +H.c.]\nonumber\\
  &+& \mu \sum_{j}(c^\dagger_{j,A}c_{j,A}+c^\dagger_{j,B}c_{j,B}).
\label{ham-1}
\ee
where $c_{j,A}$ (or $c_{j,B}$) is fermionic annihilation operator on site $A$ (or $B$) of the $jth$ cell, $A$ and $B$ are the sublattice indices, $t$ denotes the transfer integral, $\mu$ is a chemical potential and $\Delta$ denotes the superconducting pairing gap which is taken to be real here. When $\Delta=0$ and $\mu=0$, the Hamiltonian reduces to the SSH model \cite{ssh}; when $\lambda=0$, the Hamiltonian reduces to the 1D kitaev model \cite{kitaev}. This model is different with the generically dimerized Kitaev model \cite{Nagaosa,Chen,Sugimoto}.
Throughout this paper, we set $t=1$ as the unit of energy and take $\mu=0$ and $|\lambda|<1$ unless otherwise stated. We set the number of cell as $L_c$, and then the length of chain is $L_s=2L_c$.

In this section, we first consider the system with PBC. To express the Hamiltonian $H$ in the momentum space, we define a four-component operator $C_k^{\dagger}=(c_{k,A}^{\dagger},c_{k,B}^{\dagger},c_{-k,A},c_{-k,B})$, then we can rewrite the Hamiltonian as $H=\frac{1}{2}\sum_kC_k^{\dagger}h(k)C_k$, where
\be
h(k)=
\begin{pmatrix}
 0 & g(k) & 0 & w(k)\\
 g(k)^{*} & 0 & -w(k)^{*} & 0\\
 0 & -w(k) & 0 & -g(k)\\
w(k)^{*} & 0 & -g(k)^{*} & 0
\end{pmatrix}
\ee
with $g(k)=-t(1+\lambda)-t(1-\lambda)e^{-2ika}$ and $w(k)=-\Delta(1-\lambda)+\Delta(1+\lambda)e^{-2ika}$. Here $k\in[0,\pi)$ \cite{explain} and we set the lattice spacing $a=1$. After diagonalizing the Hamiltonian, we can obtain the eigenvalues
\begin{equation}
 E(k)=\pm\sqrt{A\pm B} ,
\label{tb7}
\end{equation}
where
\begin{equation}
 A=2(\Delta^2+t^2)(1+\lambda^2)-2(\Delta^2-t^2)(1-\lambda^2)\cos(2k)
\label{tb8}
\end{equation}
and
\begin{equation}
 B=|8t\Delta\lambda \cos(2k)|.
\label{tb9}
\end{equation}
It is easy to see that the gap close point is at $t=\pm\Delta\lambda$ for $k=0$ and at $\Delta=\pm t\lambda$ for $k=\frac{\pi}{2}$. We will show that these gap-closing conditions are consistent with the phase translation points. Before verifying this, we discuss the symmetry of this system.

One can easily confirm that the system is time-reversal symmetric and fulfills $Th(k)T^{-1}=h(-k)$, where the time-reversal operator is defined by $T=K$ that takes the complex conjugate. Moreover, the system has the sublattice symmetry that $C_1h(k)C_1^{-1}=-h(k)$, where the sublattice symmetry operator is defined as
\be
C_1=
\begin{pmatrix}
 1 & 0 & 0 & 0\\
 0 & -1 & 0 & 0\\
 0 & 0 & 1 & 0\\
 0 & 0 & 0 & -1
\end{pmatrix} .
\ee
Since we have $T^2=1$ and $C_1^2=1$, the topological class of this system belongs to the BDI case \cite{Ludwig}.

\begin{figure}
\includegraphics[height=70mm,width=80mm]{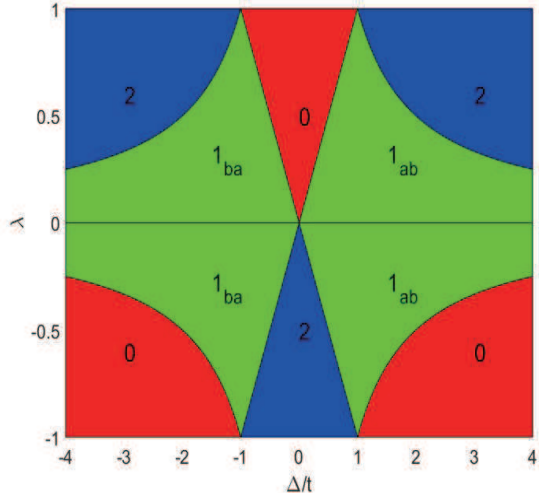}
\caption{\label{01}
  (Color online) The phase diagram obtained by using the Eq.(\ref{tb12}) and and we denote $N_{1}$ in this figure.  The horizontal axis is $\Delta/t$ and the vertical axis is $\lambda$. We can also obtain this phase diagram by using the edge correlation functions. In the thermodynamic limit, $0$ corresponds to $G_{1L}^{(1)}=0$ and $G_{1L}^{(2)}=0$, $1_{ab}$ corresponds to $G_{1L}^{(1)}\neq 0$ and $G_{1L}^{(2)}=0$, $1_{ba}$ corresponds to $G_{1L}^{(1)}=0$ and $G_{1L}^{(2)}\neq 0$ and $2$ corresponds to $G_{1L}^{(1)}\neq 0$ and $G_{1L}^{(2)}\neq 0$.}
\end{figure}

The topological number of this system can be defined as
\begin{equation}
 N_1=Tr\int_{-\frac{\pi}{2}}^{\frac{\pi}{2}}\frac{dk}{4\pi i}C_1G^{-1}\partial_kG
\label{tb10}
\end{equation}
where $G(k)=-h(k)^{-1}$ is the Green's function at zero energy \cite{Nagaosa,Gurarie,Gurarie2,Cayssol}. We then introduce a unitary transformation
\be
U_1=
\begin{pmatrix}
 1 & 0 & 0 & 0\\
 0 & 0 & 1 & 0\\
 0 & 1 & 0 & 0\\
 0 & 0 & 0 & 1
\end{pmatrix},
\ee
we have
\be
U_1h(k)U_1^{\dagger}=
\begin{pmatrix}
 0 & M_1\\
 M_1^{\dagger} & 0
\end{pmatrix}
\ee
with
\be
M_1=
\begin{pmatrix}
 g(k) & w(k)\\
 -w(k) & -g(k)\\
\end{pmatrix},
\ee
where $g(k)$ and $w(k)$ are defined earlier. Then we have \cite{Nagaosa,Chen}
\begin{eqnarray}
 N_1 &=& -Tr\int_{-\frac{\pi}{2}}^{\frac{\pi}{2}}\frac{dk}{2\pi i}M_1^{-1}\partial_kM_1 \nonumber \\
   &=& -\int_{-\frac{\pi}{2}}^{\frac{\pi}{2}}\frac{dk}{2\pi i}\partial_kln[(w-g)(w+g)] ,
 \label{tb11}
\end{eqnarray}
which gives rise to an explicit expression of the topological number
\begin{equation}
\begin{aligned}
 N_1=\Theta(|t(1-\lambda)-\Delta(1+\lambda)|-|\Delta(1-\lambda)+t(1+\lambda)|)\\
 +\Theta(|\Delta(1+\lambda)+t(1-\lambda)|-|t(1+\lambda)-\Delta(1-\lambda)|),
 \label{tb12}
\end{aligned}
\end{equation}
where $\Theta$ is the step function. Considering $|\lambda|<1$, we can obtain $N_1=\Theta(\Delta\lambda-t)+\Theta(\Delta-t\lambda)$ for $\lambda>0$ and $\Delta>0$, $N_1=\Theta(-\Delta-t\lambda)+\Theta(t+\Delta\lambda)$ for $\lambda<0$ and $\Delta>0$, $N_1=\Theta(t-\Delta\lambda)+\Theta(\Delta-t\lambda)$ for $\lambda<0$ and $\Delta<0$, and $N_1=\Theta(-\Delta-t\lambda)+\Theta(-t-\Delta\lambda)$ for $\lambda>0$ and $\Delta<0$ from Eq.(\ref{tb12}). These phase translation points are consistent with the above gap-closing conditions. Then we can obtain the phase diagram as shown in Fig.~\ref{01}. There exist three phases: (i) $|\Delta|>|t/\lambda|$ for $\lambda<0$ and $|\Delta|<t\lambda$ for $\lambda>0$, where $N_1=0$ (the trivial phase) as shown in the red region; (ii) $|t\lambda|<|\Delta|<|t/\lambda|$, where $N_1=1$ (the TSC phase) as shown in the green region; (iii) $|\Delta|>t/\lambda$ for $\lambda>0$ and $|\Delta|<|t\lambda|$ for $\lambda<0$, where $N_1=2$ (the SSH-like topological phase) as shown in the blue region of Fig.~\ref{01}.

One interesting phenomenon from Fig.~\ref{01} is that there also exist the SSH-like topological phase when $\lambda> 0$, where the hopping strength between neighbor lattice points are ``strong-weak-strong-weak-$\cdots$", which corresponds to the trivial case of the SSH model if the superconductor pairing term is omitted. When fixing $\lambda$, i.e., fixing the hopping strength of this system, we can change this system from SSH-like topological phase to the trivial phase by changing the superconductor pairing strength, which is different with the generically dimerized Kitaev model \cite{Nagaosa}. To understand these phases in the phase diagram intuitively, we investigate these phases and reconsider the phase transitions under OBC.
If taking OBC, we can't obtain the topological number. Recent work by Miao et al. \cite{Miao1,Miao2} proposed an edge correlation function to study the phase transition between TSC phase and trivial phase. Here we extend this method to introducing two edge correlations and we find the two edge correlation functions can completely characterize the phase transitions. This method gives not only a clear physical picture but also more information about the TSC phase.

\section{characterization of topological phase transition via edge correlation functions}
\label{correlation}
\subsection{Exact diagonalization}
In this section, we shall consider the system under OBC. To calculate the edge correlation functions analytically, we shall introduce the Majorana fermion representation  firstly. One fermion operator can be split into two Majorana fermion operators
\begin{subequations}
\begin{eqnarray}
c_{j,\eta} & = & \frac{1}{2}\left(\gamma_{j,\eta}^{a}+i\gamma_{j,\eta}^{b}\right),\\
c_{j,\eta}^{\dagger} & = & \frac{1}{2}\left(\gamma_{j,\eta}^{a}-i\gamma_{j,\eta}^{b}\right).
\end{eqnarray}
\end{subequations}
where $\eta=A, B$. The Majorana fermion operators fulfill
\begin{equation}
\left(\gamma_{j,\eta}^{\beta}\right)^{\dagger}=\gamma_{j,\eta}^{\beta},
\end{equation}
where $\beta=a,b$, and
\begin{equation}
\left\{ \gamma_{j,\eta}^{\beta},\gamma_{l,\eta'}^{\beta'}\right\} =2\delta_{\beta \beta'}\delta_{jl}\delta_{\eta \eta'},
\end{equation}
where $\beta'=a, b$ and $\eta'=A, B$. By using the Majorana fermion operators, the Hamiltonian
of this system can be written as
\begin{eqnarray}
&H=\frac{i}{2}&\sum_{j} \{-[t(1+\lambda)+\Delta(1-\lambda)]\gamma_{j,B}^{a}\gamma_{j,A}^{b} \nonumber\\
&&-[t(1+\lambda)-\Delta(1-\lambda)]\gamma_{j,A}^{a}\gamma_{j,B}^{b} \nonumber\\
&&-[t(1-\lambda)+\Delta(1+\lambda)]\gamma_{j+1,A}^{a}\gamma_{j,B}^{b} \nonumber\\
&&-[t(1-\lambda)-\Delta(1+\lambda)]\gamma_{j,B}^{a}\gamma_{j+1,A}^{b}.
\label{ham-2}
\end{eqnarray}
The above Hamiltonian can be simplified as
\begin{eqnarray}
H & = & \frac{i}{2}\sum_{j,l=1}^{L_s}\gamma_{j}^{a}B_{jl}\gamma_{l}^{b},
\end{eqnarray}
where for convenience we have set $\gamma_{j,A}^{\beta}=\gamma_{2j-1}^{\beta}$ and $\gamma_{j,B}^{\beta}=\gamma_{2j}^{\beta}$. Here $B$ is a $L_s\times L_s$ real matrix with $B_{2j,2j-1}=-[t(1+\lambda)+\Delta(1-\lambda)]$, $B_{2j-1,2j}=-[t(1+\lambda)-\Delta(1-\lambda)]$, $B_{2j+1,2j}=-[t(1-\lambda)+\Delta(1+\lambda)]$ and $B_{2j,2j+1}=-[t(1-\lambda)-\Delta(1+\lambda)]$.

By using the SVD, i.e., $B=U\Lambda V^T$, where $\Lambda$ is a real diagonal matrix and it gives rise to the singular values of $B$, $U$ and $V$ are two real orthogonal matrices \cite{Katsura,Lieb}, which transform the Majorana fermion operators as $\gamma_{k}^{a}=\sum_{j=1}^{L_s}U_{jk}\gamma_j^{a}$ and $\gamma_{k}^{b}=\sum_{j=1}^{L_s}V_{jk}\gamma_j^{b}$ and they satisfy $UU^{T}=VV^{T}=1$,  the Hamiltonian $H$ can be diagonalized as follows,
\begin{eqnarray}
H & = & \frac{i}{2}\sum_{k}\gamma_{k}^{a}\Lambda_{k}\gamma_{k}^{b}\nonumber \\
 & = & \sum_{k}\Lambda_{k}\left(c_{k}^{\dagger}c_{k}-\frac{1}{2}\right),
\end{eqnarray}
where $\Lambda_{k}$ are non-negative, $\gamma_k^{a}$ and $\gamma_{k}^{b}$ satisfy the relations of the Majorana fermion operators $\left(\gamma_{k}^{\beta}\right)^{\dagger}=\gamma_{k}^{\beta}$ and $\left\{ \gamma_{k}^{\beta},\gamma_{k'}^{\beta'}\right\} =2\delta_{k k'}\delta_{\beta \beta'}$, and
 $c_{k}  =  \frac{1}{2}\left(\gamma_{k}^{a}+i\gamma_{k}^{b}\right)$ and $c_{k}^{\dagger}  = \frac{1}{2}\left(\gamma_{k}^{a}-i\gamma_{k}^{b}\right)$
are fermion operators.

One can easily verify that the matrices $U, V, BB^{T}$ and $B^{T}B$ satisfy $U^{T}BB^{T}U=\Lambda^2$ and $V^{T}B^{T}BV=\Lambda^2$. The non-negative square roots of the eigenvalues of $BB^{T}$ give rise to singular values $\Lambda_{k}$ \cite{Miao1,Katsura,Lieb}. There exist two singular values
\begin{widetext}
\begin{eqnarray}
\Lambda_{k^{I}}&=&\sqrt{2(t^2+\Delta^2)(1+\lambda^2)+2[(t^2-\Delta^2)(1-\lambda^2)+4\Delta t\lambda]\cos(2k^{I})}, \nonumber\\
\Lambda_{k^{II}}&=&\sqrt{2(t^2+\Delta^2)(1+\lambda^2)+2[(t^2-\Delta^2)(1-\lambda^2)-4\Delta t\lambda]\cos(2k^{II})}.
\end{eqnarray}
\end{widetext}
We can see that $\Lambda_{k^{I}}=0$ when $\frac{\Delta}{t}=-\frac{1}{\lambda}$ and $k^{I}=0$ or when $\frac{\Delta}{t}=\lambda$ and $k^{I}=\frac{\pi}{2}$, and $\Lambda_{k^{II}}=0$ when $\frac{\Delta}{t}=\frac{1}{\lambda}$ and $k^{II}=0$ or when $\frac{\Delta}{t}=-\lambda$ and $k^{II}=\frac{\pi}{2}$. It is clear that the gap close points are at $\frac{\Delta}{t}=\pm \frac{1}{\lambda}$ and $\frac{\Delta}{t}=\pm \lambda$.

For $\Lambda_{k^{I}}$ and $\Lambda_{k^{II}}$, the corresponding $U$ and $V$ \cite{Miao1,Lieb} are
\begin{subequations}\label{eq:UVkII}
\begin{eqnarray}
U_{jk^{I}} & = & \begin{cases}
A_{k^{I}}\sin k^{I}\left(L_s+1-j\right), & j=odd,\\
0, & j=even,
\end{cases}\\
V_{jk^{I}} & = & \begin{cases}
0, & j=odd,\\
-A_{k^{I}}\delta_{k^{I}}\sin k^{I}j, & j=even.
\end{cases}
\end{eqnarray}
\end{subequations}
and
\begin{subequations}\label{eq:UVkI}
\begin{eqnarray}
U_{jk^{II}} & = & \begin{cases}
0, & j = odd,\\
A_{k^{II}}\sin k^{II}j, & j = even,
\end{cases}\\
V_{jk^{II}} & = & \begin{cases}
-A_{k^{II}}\delta_{k^{II}}\sin k^{II}\left(L_s+1-j\right), & j=odd,\\
0, & j = even.
\end{cases}
\end{eqnarray}
\end{subequations}
respectively. Here the normalization factors are
\begin{equation}
A_{k}=2\left[L_s+1-\frac{\sin2k\left(L_s+1\right)}{\sin2k}\right]^{-1/2},
\end{equation}
and
\begin{equation}
\delta_{k}=sgn[\frac{\cos k}{\cos k(L_s+1)}].
\end{equation}
The wave vector $k^I$'s are determined by
\begin{eqnarray}
\frac{\sin k^{I}\left(L_s+2\right)}{\sin k^{I}L_s}&=&-\frac{t(1-\lambda)+\Delta(1+\lambda)}{t(1+\lambda)-\Delta(1-\lambda)},
\end{eqnarray}
and $k^{II}$'s are given by
\begin{eqnarray}
\frac{\sin k^{II}\left(L_s+2\right)}{\sin k^{II}L_s}&=& -\frac{t(1-\lambda)-\Delta(1+\lambda)}{t(1+\lambda)+\Delta(1-\lambda)}.
\end{eqnarray}

\begin{figure}
\includegraphics[height=70mm,width=80mm]{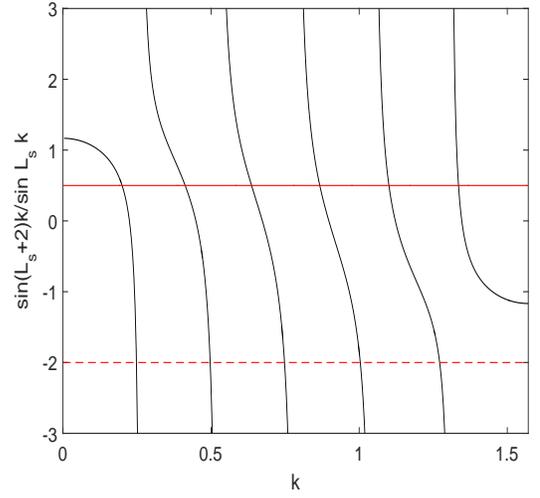}
\caption{\label{02}
  (Color online) $Sin(L_s+2)k/sin L_sk$ (black solid curve) versus $k$ for $L_s=12$, $-\frac{t(1-\lambda)+\Delta(1+\lambda)}{t(1+\lambda)-\Delta(1-\lambda)}=-2$ (red dotted line) and $-\frac{t(1-\lambda)-\Delta(1+\lambda)}{t(1+\lambda)+\Delta(1-\lambda)}=0.5$ (red solid line) for $\lambda=0.5$ and $t=\Delta=1$.}
\end{figure}
We represent the functions $\frac{sin(L_s+2)k}{sinL_sk}$ in Fig.~\ref{02} for $L_s=12$. Actually, when $k\rightarrow 0$, $\frac{sin(L_s+2)k}{sinL_sk}\rightarrow \frac{L_s+2}{L_s}$, which becomes $1$ in the thermodynamic limit $L_s\rightarrow \infty$. When $k\rightarrow \frac{\pi}{2}$, $\frac{sin(L_s+2)k}{sinL_sk}\rightarrow -1$. Therefore if $|\frac{t(1-\lambda)+\Delta(1+\lambda)}{t(1+\lambda)-\Delta(1-\lambda)}|>1$ as the red dotted line in Fig.~\ref{02}, besides $\frac{L_s}{2}$ real $k^{II}$'s and $\frac{L_s}{2}-1$ real $k^{I}$'s, there exists a single complex $k_0^{I}$,
\begin{equation}
k_{0}^{I}=\frac{\pi}{2}+iv,
\end{equation}
where $v$ fulfills
\begin{equation}
\frac{\sinh v\left(L_s+2\right)}{\sinh vL_s}=\frac{t(1-\lambda)+\Delta(1+\lambda)}{t(1+\lambda)-\Delta(1-\lambda)}.
\end{equation}
For this $k_0^{I}$ mode we have \cite{Lieb}
\begin{subequations}\label{eq:UVk0II}
\begin{eqnarray}
U_{jk_{0}^{I}} & = & \begin{cases}
A_{k_{0}^{I}}\left(-1\right)^{\frac{L_s+1-j}{2}}\sinh v\left(L_s+1-j\right) & j=odd,\\
0 & j=even,
\end{cases}\\
V_{jk_{0}^{I}} & = & \begin{cases}
0 & j=odd,\\
-A_{k_{0}^{I}}\left(-1\right)^{-\frac{L_s-j}{2}}\sinh vj & j=even.
\end{cases}
\end{eqnarray}
\end{subequations}
Then the normalization factor becomes
\begin{equation}
A_{k_{0}^{I}}=2e^{- v L_s}\left(1-e^{-4v}\right)^{1/2},
\end{equation}
and the corresponding singular value is
\begin{equation}
\Lambda_{k_{0}^{I}}\approx (1-|\frac{t(1+\lambda)-\Delta(1-\lambda)}{t(1-\lambda)+\Delta(1+\lambda)}|)|\frac{t(1+\lambda)-\Delta(1-\lambda)}{t(1-\lambda)+\Delta(1+\lambda)}|^{L_s/2}.
\end{equation}

When $|\frac{t(1-\lambda)-\Delta(1+\lambda)}{t(1+\lambda)+\Delta(1-\lambda)}|>1$, besides $\frac{L_s}{2}$ real $k^{I}$'s and $\frac{L_s}{2}-1$ real $k^{II}$'s, there exists a complex $k_0^{II}$ that $k_0^{II}=\frac{\pi}{2}+iv^{'}$, which is determined by
\begin{equation}
\frac{\sinh v^{'}\left(L_s+2\right)}{\sinh v^{'}L_s}=\frac{t(1-\lambda)-\Delta(1+\lambda)}{t(1+\lambda)+\Delta(1-\lambda)}.
\end{equation}
In a similar way, we can obtain the corresponding $U_{jk_{0}^{II}}$, $V_{jk_{0}^{II}}$ and $A_{k_0^{II}}$, which has the same expression as  $A_{k_0^{I}}$ and the corresponding singular value
\begin{equation}
\Lambda_{k_{0}^{II}}\approx (1-|\frac{t(1+\lambda)+\Delta(1-\lambda)}{t(1-\lambda)-\Delta(1+\lambda)}|)|\frac{t(1+\lambda)+\Delta(1-\lambda)}{t(1-\lambda)-\Delta(1+\lambda)}|^{L_s/2}.
\end{equation}

If $|\frac{t(1-\lambda)+\Delta(1+\lambda)}{t(1+\lambda)-\Delta(1-\lambda)}|>1$ and $|\frac{t(1-\lambda)-\Delta(1+\lambda)}{t(1+\lambda)+\Delta(1-\lambda)}|>1$, there exist two complex $k_0^{I}$ and $k_0^{II}$ besides $\frac{L_s}{2}-1$ real $k^{I}$'s and $\frac{L_s}{2}-1$ real $k^{II}$'s.
In the thermodynamic limit, $\Lambda_{k_{0}^{I}}\rightarrow 0$ and $\Lambda_{k_{0}^{II}}\rightarrow 0$, which indicates both the $k_{0}^{I}$ mode and $k_{0}^{II}$ mode giving rise to the zero mode under OBC.

\subsection{Edge correlation functions}
To study the topological phase transitions of this model, we define two edge correlation functions
\begin{equation}
G_{1L}^{(1)}=\left\langle 0\right|i\gamma_{1}^{a}\gamma_{L_s}^{b}\left|0\right\rangle.
\end{equation}
and
\begin{equation}
G_{1L}^{(2)}=\left\langle 0\right|i\gamma_{1}^{b}\gamma_{L_s}^{a}\left|0\right\rangle.
\end{equation}
where $|0\rangle$ is the ground state and its corresponding energy value is $-\frac{1}{2}\sum_{k^{I},k^{II}}(\Lambda_{k^{I}}+\Lambda_{k^{II}})$.

If $G_{1L}^{(1)}\rightarrow 0$ and $G_{1L}^{(2)}\rightarrow 0$ in the thermodynamic limit, the system is at a trivial phase. If $G_{1L}^{(1)}\rightarrow 0$ and $G_{1L}^{(2)}$ tends to a finite value or $G_{1L}^{(2)}\rightarrow 0$ and $G_{1L}^{(1)}$ tends to a finite value in the thermodynamic limit, it implies that there exists a Majorana fermion at the each end of this chain and the system is at a TSC phase. If both $G_{1L}^{(1)}$ and $G_{1L}^{(2)}$ approach to finite values in the thermodynamic limit, it indicates that there exists a Dirac fermion at the each end of this chain and the system is at the SSH-like topological phase.

The edge correlation functions $G_{1L}^{(1)}$ and $G_{1L}^{(2)}$ can be given  by
\begin{equation}
G_{1L}^{(1)}=\left\langle 0\right|i\gamma_{1}^{a}\gamma_{L_s}^{b}\left|0\right\rangle=-\sum_{k}U_{1k}V_{L_sk}.
\end{equation}
and
\begin{equation}
G_{1L}^{(2)}=\left\langle 0\right|i\gamma_{1}^{b}\gamma_{L_s}^{a}\left|0\right\rangle=-\sum_{k}V_{1k}U_{L_sk}.
\end{equation}
Since the ground state is composed of the $k^{I}$ mode and $k^{II}$ mode, we need investigate the effect of both $U_I$, $V_I$ and $U_{II}$, $U_{II}$ on the two edge correlation functions. Because $U_{1k^{II}}=V_{L_sk^{II}}=0$ and $U_{L_sk^{I}}=V_{1k^{I}}=0$, we have $G_{1L}^{(1)}=-\sum_{k^{I}}U_{1k^{I}}V_{L_sk^{I}}$ and  $G_{1L}^{(2)}=-\sum_{k^{II}}V_{1k^{II}}U_{L_sk^{II}}$.
When $|\frac{t(1-\lambda)+\Delta(1+\lambda)}{t(1+\lambda)-\Delta(1-\lambda)}|<1$ and $|\frac{t(1-\lambda)-\Delta(1+\lambda)}{t(1+\lambda)+\Delta(1-\lambda)}|<1$, there isn't zero mode in this system. Then we have
\begin{eqnarray}
G_{1L}^{(1)}=\sum_{k^{I}}A_{k^{I}}^{2}\delta_{k^{I}}\sin^{2}k^{I}L_s.
\end{eqnarray}
which can be proven to be of order of $O\left(1/L_s\right)$ by following Lieb et al. \cite{Lieb}, and
\begin{eqnarray}
G_{1L}^{(2)}=\sum_{k^{II}}A_{k^{II}}^{2}\delta_{k^{II}}\sin^{2}k^{II}L_s= O\left(1/L_s\right).
\end{eqnarray}
which means that there exists no Majorana fermion at the end of the chain and the system is topologically trivial.

\begin{figure}
\includegraphics[height=45mm,width=80mm]{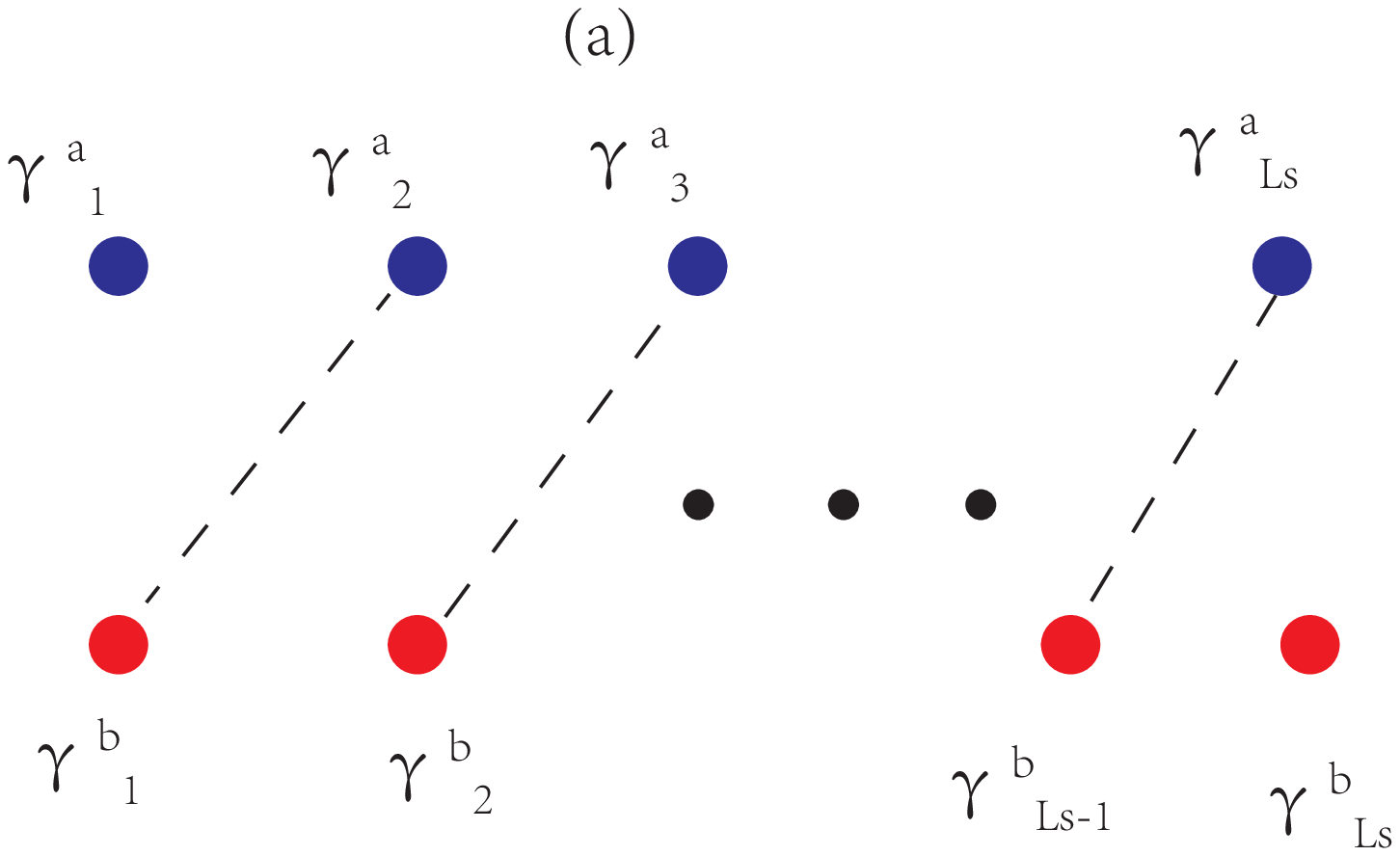}\\
\includegraphics[height=45mm,width=80mm]{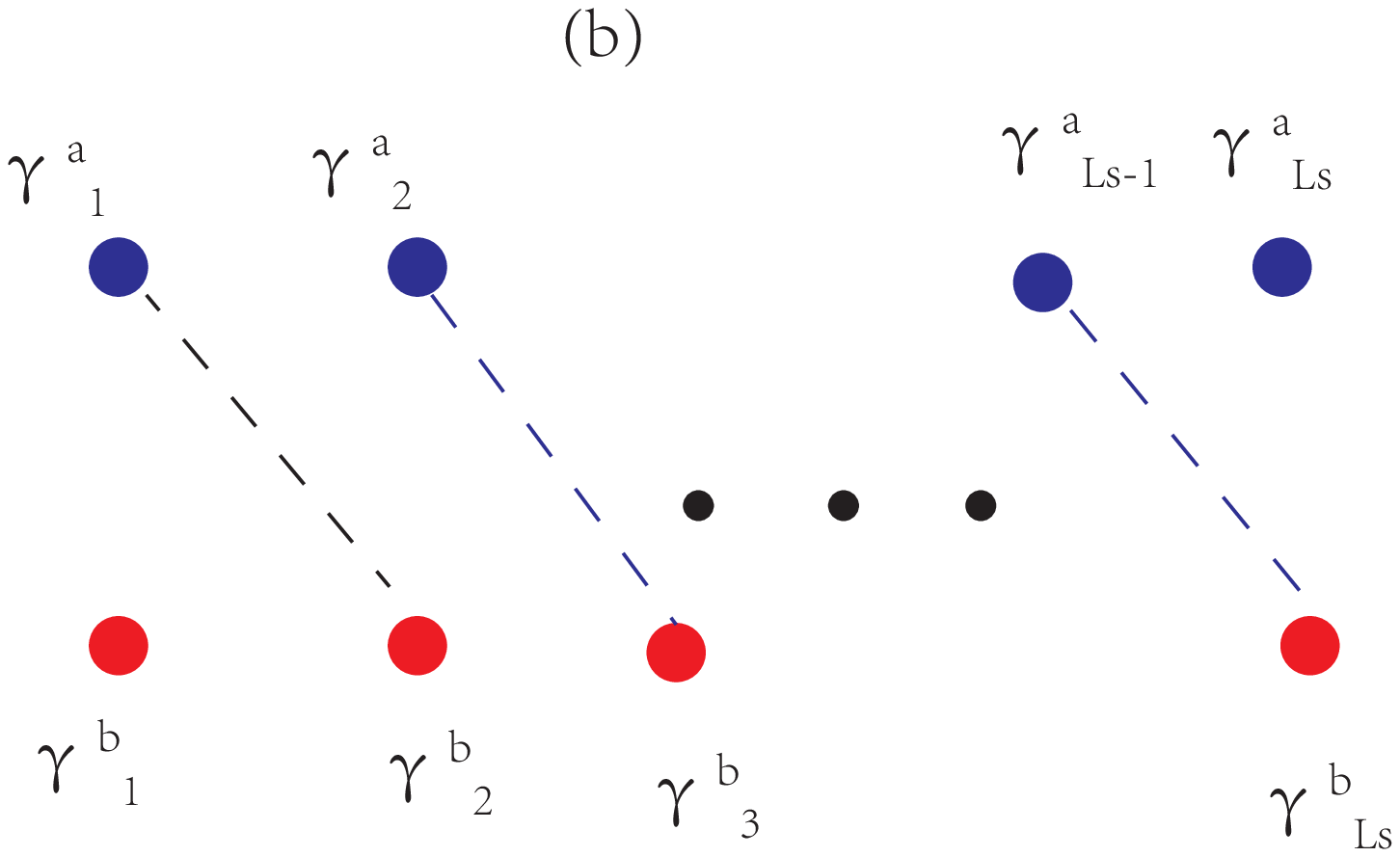}
\caption{\label{03}
  (Color online) Schematic cartoon picture for the coupling pattern of the Hamiltonian for the (a) $G_{1L}^{(1)}\neq 0$ and $G_{1L}^{(2)}= 0$ and (b) $G_{1L}^{(1)}= 0$ and $G_{1L}^{(2)}\neq 0$ in the thermodynamic limit.}
\end{figure}

When $|\frac{t(1-\lambda)+\Delta(1+\lambda)}{t(1+\lambda)-\Delta(1-\lambda)}|>1$ and $|\frac{t(1-\lambda)-\Delta(1+\lambda)}{t(1+\lambda)+\Delta(1-\lambda)}|<1$, there exists a zero mode in the system, then
the edge correlation function is
\begin{eqnarray}
G_{1L}^{(1)} & = & \left\langle 0\right| i\gamma_{1}^{a}\gamma_{L_s}^{b} \left|0\right\rangle = - U_{1k_{0}^{I}}V_{L_sk_{0}^{I}}-\sum_{k^{I}}U_{1k^{I}}V_{L_sk^{I}}\nonumber \\
 & = & \left(-1\right)^{L_s/2}A_{k_{0}^{I}}^{2}\sinh^{2} v L_s+\sum_{k^{I}}A_{k^{I}}^{2}\delta_{k^{I}}\sin^{2}k^{I}L_s\nonumber \\
 & = & \left(-1\right)^{L_s/2}\left[1-\left(\frac{t(1+\lambda)-\Delta(1-\lambda)}{t(1-\lambda)+\Delta(1+\lambda)}\right)^{2}\right]+O\left(1/L_s\right)
 \notag
\end{eqnarray}
and $G_{1L}^{(2)}=O(1/L_s)$. For this case, there exists a Majorana fermion at each end of the chain, which corresponds to the TSC phase, as schematically shown in Fig.~\ref{03}(a).
When $|\frac{t(1-\lambda)+\Delta(1+\lambda)}{t(1+\lambda)-\Delta(1-\lambda)}|<1$ and $|\frac{t(1-\lambda)-\Delta(1+\lambda)}{t(1+\lambda)+\Delta(1-\lambda)}|>1$,
we can obtain
\begin{eqnarray}
G_{1L}^{(2)} & = & \left\langle 0\right|i\gamma_{1}^{b}\gamma_{L_s}^{a}\left|0\right\rangle \nonumber \\
 & = & \left(-1\right)^{L_s/2}\left[1-\left(\frac{t(1+\lambda)+\Delta(1-\lambda)}{t(1-\lambda)-\Delta(1+\lambda)}\right)^{2}\right]+O\left(1/L_s\right)
 \notag
\end{eqnarray}
and $G_{1L}^{(2)}=O(1/L_s)$. This case also corresponds to the TSC phase, as shown in Fig.~\ref{03}(b).

When $|\frac{t(1-\lambda)+\Delta(1+\lambda)}{t(1+\lambda)-\Delta(1-\lambda)}|>1$ and $|\frac{t(1-\lambda)-\Delta(1+\lambda)}{t(1+\lambda)+\Delta(1-\lambda)}|>1$, there exist two degenerate zero modes in the system. One can easily verify that
$G_{1L}^{(1)}=\left(-1\right)^{L_s/2}\left[1-\left(\frac{t(1+\lambda)-\Delta(1-\lambda)}{t(1-\lambda)+\Delta(1+\lambda)}\right)^{2}\right]$ and $G_{1L}^{(2)}=\left(-1\right)^{L_s/2}\left[1-\left(\frac{t(1+\lambda)+\Delta(1-\lambda)}{t(1-\lambda)-\Delta(1+\lambda)}\right)^{2}\right]$ in the thermodynamic limit.
For this case, there exist two Majorana fermions at each end of the chain, i.e., there exists a Dirac fermion at each end of the chain, which corresponds to the SSH-like topological phase.

From the above discussions, we can obtain the phase diagram as shown in Fig.~\ref{01}, where $0$ represents the trivial phase corresponding  to $G_{1L}^{(1)}=0$ and $G_{1L}^{(2)}=0$ in the thermodynamic limit, $1_{ab}$ represents the TSC phase corresponding to $G_{1L}^{(1)}\neq 0$ and $G_{1L}^{(2)}=0$, $1_{ba}$ corresponding to $G_{1L}^{(1)}=0$ and $G_{1L}^{(2)}\neq 0$,  and $2$ represents the SSH-like topological phase corresponding to $G_{1L}^{(1)}\neq 0$ and $G_{1L}^{(2)}\neq 0$. Our results are consistent with the analytical results by calculating the topological number $N_1$.

\begin{figure}
\includegraphics[height=50mm,width=80mm]{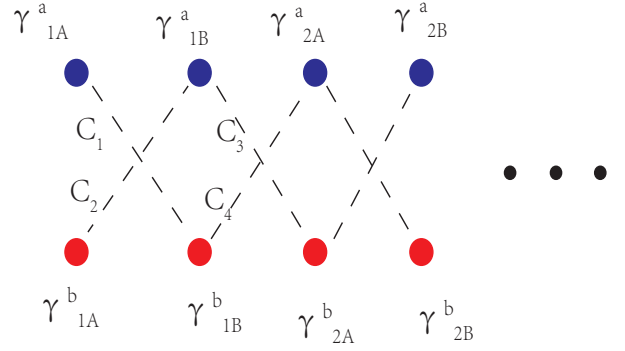}
\caption{\label{04}
Schematic cartoon picture for the coupling pattern of the Hamiltonian (\ref{ham-3}). Here $C_1, C_2, C_3$ and $C_4$ are the coupling constants defined in the text.}
\end{figure}

For the phase of  $N_1=2$ in the region of $\lambda<0$, the hopping strength between neighbor sites is ``weak-strong-weak-strong-$\cdots$", which is consistent with the topological SSH model, corresponding to the SSH-like topological phase. However, for the phase of  $N_1=2$ in the region of $\lambda>0$, the hopping strength between neighbor site becomes ``strong-weak-strong-weak-$\cdots$" and the superconductor pairing of neighbor lattice points becomes ``weak-strong-weak-strong-$\cdots$". To illustrate how the SSH-like topological phase emerges even for $\lambda>0$, we rewrite the Hamiltonian (\ref{ham-2}) as
\begin{equation}
\begin{aligned}
&H = -\frac{i}{2} \sum_{j} [C_1\gamma_{j,A}^{a}\gamma_{j,B}^{b}+C_2\gamma_{j,A}^{b}\gamma_{j,B}^{a}+C_3\gamma_{j,B}^{a}\gamma_{j+1,A}^{b}\\
&+C_4\gamma_{j,B}^{b}\gamma_{j+1,A}^{a}],
\label{ham-3}
\end{aligned}
\end{equation}
where $C_1=t(1+\lambda)-\Delta(1-\lambda)$, $C_2=-t(1+\lambda)-\Delta(1-\lambda)$, $C_3=t(1-\lambda)-\Delta(1+\lambda)$ and $C_4=-t(1-\lambda)-\Delta(1+\lambda)$. We show the illustration of the coupling pattern of Majorana fermions for the Hamiltonian in Fig.~\ref{04}. For the case of $N_1=2$, we have $|\frac{t(1-\lambda)+\Delta(1+\lambda)}{t(1+\lambda)-\Delta(1-\lambda)}|>1$ and $|\frac{t(1-\lambda)-\Delta(1+\lambda)}{t(1+\lambda)+\Delta(1-\lambda)}|>1$, i.e., $|C_4|>|C_1|$ and $|C_3|>|C_2|$, and therefore there exists a Dirac fermion at the each end of the chain, similar to the case of the SSH model. Similarly, for the case of $N_1=0$, we have $|\frac{t(1-\lambda)+\Delta(1+\lambda)}{t(1+\lambda)-\Delta(1-\lambda)}|<1$ and $|\frac{t(1-\lambda)-\Delta(1+\lambda)}{t(1+\lambda)+\Delta(1-\lambda)}|<1$, i.e., $|C_4|<|C_1|$ and $|C_3|<|C_2|$, and no edge state occurs. Furthermore, for the case of $N_1=1$, we have $|C_4|>|C_1|$ and $|C_3|<|C_2|$, or  $|C_4|<|C_1|$ and $|C_3|>|C_2|$, corresponding to the coupling pattern $1_{ab}$ and $1_{ba}$, respectively.

\section{Survival probability of an edge Majorana under sudden quenching of parameters}
\label{dynamics}
\begin{figure}
\includegraphics[height=90mm,width=90mm]{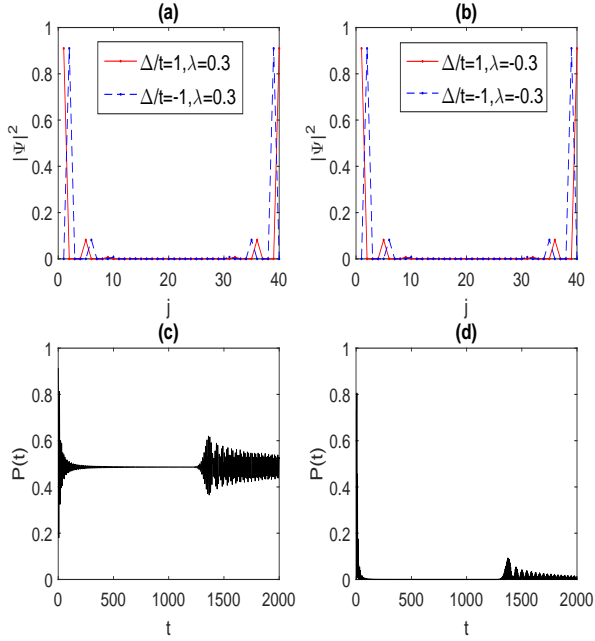}
\caption{\label{05}
 The distributions of two zero-energy modes of the open chain with $L_s=20$, (a) $\Delta/t=1,\lambda=0.3$ (red line), $\Delta/t=-1,\lambda=0.3$ (black line), (b) $\Delta/t=1,\lambda=-0.3$ (red line) and $\Delta/t=-1,\lambda=-0.3$ (black line). The survival probability of an edge Majorana for an open chain with $L_s=200$ as a function of time for the quench from (c) $\Delta/t=1, \lambda=0.3 \to \Delta/t=1, \lambda=-0.3$, (d) $\Delta/t=1, \lambda=0.3 \to \Delta/t=-1, \lambda=0.3$.}
\end{figure}

The two edge correlation functions give rise to two different coupling cases for the TSC phase labeled by $1_{ab}$ and $1_{ba}$, respectively. In this section we will investigate the survival probability of an edge Majorana when quenching between the two different coupling cases and quenching in the same coupling case. To study the quench dynamics of the system, we represent the Hamiltonian (\ref{ham-2}) as
\begin{equation}
H = \frac{i}{4}\sum_{j,k=1}^{2L_s}\gamma_{j}A_{jk}\gamma_{k}.
\label{ham-4}
\end{equation}
Here we set $\gamma_{j,A}^{a}=\gamma_{4j-3}$, $\gamma_{j,A}^{b}=\gamma_{4j-2}$, $\gamma_{j,B}^{a}=\gamma_{4j-1}$, $\gamma_{j,B}^{b}=\gamma_{4j}$ and $A$ is a $2L_s \times 2L_s$ matrix.
After diagonalizing the Hamiltonian (\ref{ham-4}) with given parameters in the TSC phase, we can obtain two zero-energy modes $E_{L_s}$ and $E_{L_s+1}$, which are the $L_s-th$ and $(L_s+1)-th$ eigenvalues with the corresponding eigenstates  $\psi_{L_s}$ and $\psi_{L_s+1}$. Fig.~\ref{05}(a) and (b) show the distribution of these two zero-energy modes, which is defined as $|\Psi_{j}|^2=|\psi_{L_s,j}|^2+|\psi_{L_s+1,j}|^2$, where $j=1,2,\dots, 2L_s$ labels the site of Majorana chain. One can see that the major distribution is at $j=1$ and $j=2L_s$ for the case of $\left\langle 0\right|i\gamma_{1}^{a}\gamma_{L_s}^{b}\left|0\right\rangle \neq 0$ and the major distribution is at $j=2$ and $j=2L_s-1$  for the case of $\left\langle 0\right|i\gamma_{1}^{b}\gamma_{L_s}^{a}\left|0\right\rangle \neq 0$.

We now perform a sudden quench of the Hamiltonian by suddenly changing the parameter from $\lambda$ to  $- \lambda$ or from $\Delta$ to $-\Delta$, and study the survival probability of zero-energy Majorana mode after a long time.  Following a sudden quench to final Hamiltonian $H_f$, the final state at time $t$ can be written as
\begin{equation}
\begin{aligned}
|\Psi(t)\rangle=e^{-iH_ft}|\Psi(0)\rangle=\sum_{n=1}^{2L_s}e^{-iE_nt}|\psi_n\rangle\langle\psi_n|\Psi(0)\rangle,
\label{final}
\end{aligned}
\end{equation}
where $|\Psi(0)\rangle$ is an initial state, $|\psi_n\rangle$ is the $n-th$ eigenstate of the final Hamiltonian and $E_n$ is the corresponding eigenvalue. We set the initial state as $|\Psi(0)\rangle=|\psi_{L_s}\rangle$ with the parameters $\Delta/t=1,\lambda=0.3$, which is an edge Majorana state. Then we can define the survival probability of this edge Majorana state as
 \begin{equation}
\begin{aligned}
 P(t)=|\langle\Psi(0)|\Psi(t)\rangle|^2,
\label{probability}
\end{aligned}
\end{equation}
which is actually the Loschmidt echo \cite{Gorin,Sun,Dutta,Heyl,Andraschko,Mazza,Jafari}.

Fig.~\ref{05}(c) shows the survival probability $P$ of an edge Majorana versus time $t$ for the quench from  $\Delta/t=1, \lambda=0.3$ to $\Delta/t=1, \lambda=-0.3$. In this case, the initial and finial Hamiltonian lays in the same region  and we see that the survival probability $P$ never approaches zero during the evolution process. In Fig.~\ref{05}(d), we present the survival probability $P$ of an edge Majorana as a function of time $t$ for the quench from $\Delta/t=1, \lambda=0.3$ to $\Delta/t=-1, \lambda=0.3$, which corresponds to the case with the initial and the finial Hamiltonian located in different regions labeled by  $1_{ab}$ and $1_{ba}$, respectively. One can see that the survival probability $P$ can reach nearby zero after a finite time evolution, which means that the message of the initial edge Majorana state completely disappears.

\section{Summary}
\label{summary}
In summary, we have investigated a modified dimerized Kitaev model analytically and demonstrated it exhibiting a rich phase diagram. There exist three different phases, i.e., the trivial, SSH-like topological and TSC phases, in different parameter regions. We identified the phase diagram by using two different methods by calculating the topological number and two introduced edge correlation functions of Majorana fermions, respectively. Comparing these two methods, we find the latter method is more intuitive and gives directly the coupling information for the edge Majorana fermions. When this two edge correlation functions both equal zero in the thermodynamic limit, the system is at the trivial phase. When they are both nonzero, the system is at the SSH-like topological phase. When one of them equals zero and the other is  nonzero, there exists a Majorana fermion at each end of the chain, i.e., the system is at a TSC phase. Furthermore, the edge correlation functions can distinguish two different coupling patterns of Majorana fermions at the TSC phase, which can also be unveiled by investigating the survival probability of an edge Majorana under sudden quenching of parameters.


\begin{acknowledgments}
The work is supported by the National Key Research and Development Program of China (2016YFA0300600), NSFC under Grants No. 11425419, No. 11374354 and No. 11174360, and the Strategic Priority Research Program (B) of the Chinese Academy of Sciences  (No. XDB07020000).
\end{acknowledgments}



\begin{thebibliography}{25}
\bibitem{Zhang} M. Z. Hasan and C. L. Kane, Rev. Mod. Phys. {\bf 82}, 3045 (2010);
X.-L. Qi and S.-C. Zhang, Rev. Mod. Phys. {\bf 83}, 1057 (2011).
\bibitem{Ludwig} A. P. Schnyder, S. Ryu, A. Furusaki, and A. W. W. Ludwig,
Phys. Rev. B {\bf 78}, 195125 (2008).
\bibitem{Ryu} C.-K. Chiu, J. C. Y. Teo, A. P. Schnyder, and S. Ryu,
Rev. Mod. Phys. {\bf 88}, 035005 (2016).
\bibitem{Tanaka} Y. Tanaka, M. Sato, and N. Nagaosa,
J. Phys. Soc. Jpn. {\bf 81}, 011013 (2012).
\bibitem{kitaev} A. Y. Kitaev, Phys. Usp. {\bf 44}, 131 (2001).
\bibitem{Kane} L. Fu and C. L. Kane,
Phys. Rev. Lett. {\bf 100}, 096407 (2008).
\bibitem{Alicea} J. Alicea, Phys. Rev. B {\bf 81}, 125318 (2010).
\bibitem{Sarma2} J. D. Sau, R. M. Lutchyn, S. Tewari, and S. Das Sarma,
Phys. Rev. Lett. {\bf 104}, 040502 (2010).
\bibitem{Fujimoto} M. Sato, Y. Takahashi, and S. Fujimoto,
Phys. Rev. B {\bf 82}, 134521 (2010).
\bibitem{Sarma3} T. D. Stanescu, R. M. Luchyn, ans S. Das Sarma,
Phys. Rev. B {\bf 84}, 144522 (2011).
\bibitem{Mourik} V. Mourik, K. Zuo, S. M. Frolov, S. R. Plissard, E. P. A. M. Bakkers,
and L. P. Kouwenhoven, Science {\bf 336}, 1003 (2012).
\bibitem{Lang} L.-J. Lang and S. Chen, Phys. Rev. B. {\bf 86}, 205135 (2012).
\bibitem{Chen2} X. Cai, L-J. Lang, S. Chen, and Y. Wang,
Phys. Rev. Lett. {\bf 110}, 176403 (2013).
\bibitem{Sen} W. DeGottardi, D. Sen, and S. Vishveshwara,
Phys. Rev. Lett. {\bf 110}, 146404 (2013).
\bibitem{Hu} J. Wang, X.-J. Liu, G. Xianlong, and H. Hu,
Phys. Rev. B {\bf 93}, 104504 (2016).
\bibitem{Wang} Y. Wang, Y. Wang, and S. Chen,
Eur. Phys. J. B {\bf 89}, 254 (2016).
\bibitem{Loss} S. Hoffman, J. Klinovaja, and D. Loss,
Phys. Rev. B {\bf 93}, 165418 (2016); J. Klinovaja, P. Stano, and D. Loss,
Phys. Rev. Lett. {\bf 109}, 236801 (2012).
\bibitem{Nagaosa} R. Wakatsuki, M. Ezawa, Y. Tanaka, and N. Nagaosa,
Phys. Rev. B {\bf 90}, 014505 (2014).
\bibitem{Chen} Q-B Zeng, S. Chen, and R. L\"{u},
Phys. Rev. B {\bf 94}, 125408 (2016).
\bibitem{Sugimoto} T. Sugimoto, M. Ohtsu, and T. Tohyama,
arXiv:1708.00982.
\bibitem{Zhou} B. Zhou and S. Q. Shen,
Phys. Rev. B {\bf 84}, 054532 (2011).
\bibitem{Kells} G. Kells, D. Meidan, and P. W. Brouwer,
Phys. Rev. B {\bf 85}, 060507 (2012);
M. T. Rieder, G. Kells, M. Duckheim, D. Meidan, and P. W. Brouwer,
Phys. Rev. B {\bf 86}, 125423 (2012).
\bibitem{Wakatsuki} R. Wakatsuki, M. Ezawa, and N. Nagaosa,
Phys. Rev. B {\bf 89}, 174514 (2014).
\bibitem{ssh} W. P. Su, J. R. Schrieffer, and A. J. Heeger,
Phys. Rev. Lett. {\bf 42}, 1698 (1979).
\bibitem{ssh2} W. P. Su, J. R. Schrieffer, and A. J. Heeger,
Phys. Rev. B {\bf 22}, 2099 (1980).
\bibitem{Patel} A. A. Patel, S. Sharma, and A. Dutta,
Eur. Phys. J. B {\bf 86}, 367 (2013).
\bibitem{Rajak} A. Rajak and A. Dutta, Phys. Rev. E {\bf 89}, 042125 (2014).
\bibitem{Sacramento} P. D. Sacramento, Phys. Rev. E {\bf 93}, 062117 (2016).
\bibitem{explain} We use the Fourier transform: $c_{k,A}=1/\sqrt{L_s}\sum_je^{-ik(2j-1)a}c_{j,A}$ and $c_{k,B}=1/\sqrt{L_s}\sum_je^{-ik2ja}c_{j,B}$.
Then we obtain the matrix $h(k)$ as shown in our paper, here $k\in[0,\pi)$.
 If one use the Fourier transform: $c_{k,\eta}=1/\sqrt{L_s}\sum_je^{-ikja}c_{j,\eta}$, where $\eta=A, B$.
Then one can obtain the same matrix $h(k)$ but with $g(k)=-t(1+\lambda)-t(1-\lambda)e^{-ika}$ and $w(k)=-\Delta(1-\lambda)+\Delta(1+\lambda)e^{-ika}$,
$k\in[0,2\pi)$ if we set the lattice spacing $a=1$. Both the two cases give the same results. We use the first case in our paper.
\bibitem{Gurarie} V. Gurarie, Phys. Rev. B {\bf 83}, 085426 (2011).
\bibitem{Gurarie2} S. R. Manmana, A. M. Essin, R. M. Noack, and V. Gurarie, Phys. Rev. B {\bf 86}, 205119 (2012).
\bibitem{Cayssol} D. Sticlet, L. Seabra, F. Pollmann, and J. Cayssol,
Phys. Rev. B {\bf 89}, 115430 (2014).
\bibitem{Miao1} J.-J. Miao, H.-K. Jin, F.-C. Zhang and Y. Zhou,
arxiv:1608.08382 (2016).
\bibitem{Miao2} J.-J. Miao, H.-K. Jin, F.-C. Zhang and Y. Zhou,
Phys. Rev. Lett. {\bf 118}, 267701 (2017).
\bibitem{Katsura} H. Katsura, D. Schuricht, and M. Takahashi, Phys. Rev. B {\bf 92}, 115137 (2015).
\bibitem{Lieb} E. Lieb, T. Schultz and D. Mattis, Ann. Phys. 16, 407 (1961).
\bibitem{Gorin} T. Gorin, T. Prosen, T. H. Seligman, and M. Znidaric, Phys. Rep. {\bf 435}, 33 (2006).
\bibitem{Sun} H. T. Quan, Z. Song, X. F. Liu, P. Zanardi, and C. P. Sun, Phys. Rev. Lett. {\bf 96}, 140604 (2006).
\bibitem{Dutta} V. Mukherjee, S. Sharma, and A. Dutta, Phys. Rev. B {\bf 86},
020301(R) (2012); S. Sharma, V. Mukherjee, and A. Dutta, Eur. Phys. J. B {\bf 85}, 143
(2012).
\bibitem{Heyl} M. Heyl, A. Polkovnikov and S. Kehrein, Phys. Rev. Lett. {\bf 110}, 135704 (2013); M. Heyl, Phys. Rev. Lett. {\bf 113}, 205701 (2014);
 M. Heyl, Phys. Rev. Lett. {\bf 115}, 140602 (2015).
\bibitem{Andraschko} F. Andraschko and J. Sirker, Phys. Rev. B 89, 125120 (2014).
\bibitem{Mazza} P. P. Mazza, J. M. St¡äephan, E. Canovi, V. Alba, M. Brockmann,
and M. Haque, J. Stat. Mech. (2016) P013104
\bibitem{Jafari} R. Jafari and H. Johannesson, Phys. Rev. Lett. {\bf 118}, 015701 (2017).
\end{thebibliography}
\end{document}